\documentclass[submission,copyright,creativecommons]{eptcs}
\providecommand{\event}{ACL2 2017} 
\usepackage{breakurl}             
\usepackage{underscore}           
\usepackage{comment}
\usepackage{fancyvrb}

\usepackage{float}
\floatstyle{boxed}
\restylefloat{figure}

\title{Proof Reduction of Fair Stuttering Refinement of Asynchronous Systems and Applications}
\author{Rob Sumners
\institute{Centaur Technology}
\email{rsumners@centtech.com}
}
\def\titlerunning{Proof Reductions for Fair Stuttering Refinement of Asynchronous Systems}
\def\authorrunning{Rob Sumners}
\begin{document}
\maketitle

\begin{abstract}
  We present a series of definitions and theorems demonstrating how to reduce
  the requirements for proving system refinements ensuring containment of fair
  stuttering runs. A primary result of the work is the ability to reduce the
  requisite proofs on runs of a system of interacting state machines to a set
  of definitions and checks on single steps of a small number of state machines
  corresponding to the intuitive notions of freedom from starvation and
  deadlock. We further refine the definitions to afford an efficient
  explicit-state checking procedure in certain finite state cases. We
  demonstrate the proof reduction on versions of the Bakery Algorithm.
\end{abstract}

\section{Introduction} \label{sec:intro}

Much of hardware and software system design focuses on how to optimize the
execution of tasks by dividing the tasks into smaller computations and then
scheduling and distributing these computations on the available resources. The
natural specification for these systems is an assurance that the systems
eventually complete the supplied tasks with results consistent with an atomic
(or as atomic as feasible) execution of the task. We refresh the notion of fair
stuttering refinements~\cite{RaySumners} as a means of codifying these
specifications -- a fair stuttering refinement between two systems ensures that
every infinite run of a lower-level system with fair selection and finite
stuttering maps to a similarly restricted infinite run of a higher-level
system. This notion of refinement can allow sequences of smaller steps in the
implementation to be mapped to single steps in the specification while
additionally requiring that every task makes progress to completion.

Many previous efforts~\cite{RaySumners} have attempted to improve the
capability of theorem provers in reasoning about refinements for distributed
and concurrent systems. Previous efforts in regards to the ACL2 theorem
prover~\cite{ACL2} focused on trying to reduce the proofs of stuttering
refinements with additional structures added to define fair selection and
ensuring progress. These efforts generally boiled down to showing that a
specification could match the step of an implementation or the implementation
stuttered and some {\tt rank} function decreased. The primary difficulty in
these proofs was defining and proving an inductive invariant (either through
ACL2 or trying to prove the invariant through some form of state
exploration). In addition, the inclusion of additional structures to track
fairness and progress as well as the resulting definition of {\tt rank}
functions proved complex. Further, the additional structures at times
obfuscated whether the specification was complete and accurate.

In this paper, we take a different tack. We assume certain characteristics of
the system we are trying to verify and leverage these characteristics in
reducing the proof obligations. In particular, we first assume that the systems
we are trying to verify are asynchronous in terms of how tasks make progress to
completion. Further, we require the system definition to split the normal
next-state transition relation into a next-state relation which only takes
forward steps and a blocking relation which defines precisely when a task is
blocked from making progress. From these assumed characteristics, we define
proof reductions which reduce the goal of proving fair stuttering refinement to
proving properties of a few task steps in relation to each other. These proof
reductions have been formally defined and mechanically proven in ACL2 and are
included in the supporting materials for this paper. In the remainder of this
paper, we will cover two stages of proof reductions, review the application of
the reductions to a version of the Bakery Algorithm. We conclude the paper with
further reductions targeting efficient automatic checks in the finite state case.

\section{Preliminaries} \label{sec:prelim}

Commonly, systems are defined by an initial state predicate: {\tt (init x)} and
a next-state relation: {\tt (next x y)}. A run of the system is then simply a
sequence of states where the first state satisfies {\tt (init x)} and each pair
of states in the sequence satisfies {\tt (next x y)}. We extend this basic
construction in a couple of ways. 

First, our goal is to reason about fair executions of a system (either as an
assumption of fair selection for which task will update next or as a guarantee
that every task makes progress). Thus, we assume that there is some set of task
identifiers recognized by a predicate {\tt (id-p k)} and add a task id
parameter to the next-state relation: {\tt (next x y k)} where this now relates
state {\tt x} to state {\tt y} for an update to the task with id {\tt k}. We
also assume only one task updates at each step of the system without any
prescribed order of task updates -- essentially, the system is asynchronous at
the level of task updates.

Second, we will find it useful to require the definition of an additional
relation {\tt (blok x k)} which returns true when the task identified by {\tt
  k} is currently blocked from making progress in state {\tt x}. Further, with
this required definition of {\tt (blok x k)}, we will also require the theorem:
{\tt (not (next x x k))} be proven and use inequality of next-states as a
marker that a task is making progress to completion.

A system is then defined by three functions: {\tt (init x)}, {\tt (next x y
  k)}, and {\tt (blok x k)}. Our final goal is to prove that the fair runs of
an implementation system map to fair runs of a specification system with an
allotment for finite stuttering and some guarantee of progress. A run of a
system is a function {\tt (run i)} which takes a natural {\tt i} and returns a
state of the system. Runs will naturally need to satisfy some constraints as
detailed in Figure~\ref{fig:run}. For a given system named {\tt sys}, the macro
{\tt (def-inf-run sys)} assumes the definition of {\tt (sys-init x)}, {\tt
  (sys-next x y k)}, {\tt (sys-blok x k)}, {\tt (sys-pick i)}, {\tt (sys-run
  i)} and generates the definitions and theorems defining the properties for
the run as in Figure~\ref{fig:run}.

\begin{figure}
\footnotesize
\begin{verbatim}
(encapsulate
 ((run  (i) t)
  (pick (i) t))
 (local (defun run  (i) ....))
 (local (defun pick (i) ....))
 
 (defun step (x y k)
   (if (or (null k)    ;; finite stutter
           (blok x k)) ;; or k is blocked in x
       (equal x y)
     (next x y k)))
 
 (defthm run-init-thm (implies (zp i)   (init (run i))))
 (defthm run-step-thm (implies (posp i) (step (run (1- i)) (run i) (pick i))))
)
\end{verbatim}
\normalsize
\caption{Definition of an infinite run in ACL2}
\label{fig:run}
\end{figure}

\begin{figure}
\footnotesize
\begin{verbatim}
(defthm fair-nat-thm (natp (fair k i)))

(defthm pick-fair-thm
  (implies (and (posp i)
                (id-p k)
                (not (equal (pick i) k)))
           (< (fair k i) (fair k (1- i)))))
\end{verbatim}
\normalsize
\caption{{\em Fair Runs}: fair task selection during a run}
\label{fig:fair}
\end{figure}

\begin{figure}
\footnotesize
\begin{verbatim}
(defthm prog-is-nat (natp (prog k i)))

(defthm run-prog-thm
  (implies (and (posp i)
                (id-p k)
                (or (not (equal (pick i) k))
                    (equal (run i) (run (1- i)))))
           (< (prog k i) (prog k (1- i)))))
\end{verbatim}
\normalsize
\caption{{\em Valid Runs}: ensuring task progress during a run}
\label{fig:prog}
\end{figure}

Of particular note, the function {\tt (step x y k)} relates states {\tt x} and
{\tt y} via {\tt (next x y k)} only if {\tt k} is not blocked in {\tt x} and we
are not stuttering (denoted by the input {\tt k} being {\tt nil}) -- (as a
note, the only requirement we place on {\tt id-p} is that {\tt (not (id-p
  nil))}). So, an infinite run is defined by two functions {\tt (run i)} which
defines the sequence of states and {\tt (pick i)} which defines the sequence of
task identifiers selected. We constrain {\tt (pick i)} to only return an {\tt
  id-p} or {\tt nil}. We can now naturally define fair selection of {\tt (pick
  i)} by positing the existence of a function {\tt (fair k i)} which returns
natural numbers and for each task id {\tt k} will strictly decrease when {\tt
  k} is not selected -- see Figure~\ref{fig:fair}. The macro {\tt
  (def-fair-pick sys id-p)} assumes the definitions of {\tt (sys-pick i)}, {\tt
  (sys-fair k i)}, and {\tt (id-p k)} and produces the theorems in
Figure~\ref{fig:fair}. We use the term {\em fair run} for an infinite run with
a fair picker.

Fair selection of task identifiers ensures that each run only has finite
stuttering and that each task gets a chance to make progress, but it does not
guarantee that tasks actually make progress. We introduce the term {\em valid
  run} for a run which is not only fair but ensures progress for each task. In
order to ensure progress, we define a function {\tt (prog k i)} similar to {\tt
  (fair k i)} but in addition to ensuring {\tt pick} eventually equals {\tt k},
we also need to ensure that a state change actually occurs. The properties in
Figure~\ref{fig:prog} ensure a {\em valid run} and the macro {\tt
  (def-valid-run sys id-p)} produces these theorems for {\tt id-p}, {\tt
  sys-run}, {\tt sys-pick}, and {\tt sys-prog}. We note that a valid run is
also a fair run and thus our notion of refinement is compositional -- but it is
better to prove that all fair runs of the implementation are valid runs and
then restrict the refinement to valid runs mapping to valid runs and reduce the
proof requirements accordingly at each step. This is straightforward from what
we present in this paper but we do not focus on it in this paper.

\section{Proof Reduction to Single System Steps} \label{sec:single}

The principle objective of fair stuttering refinement is to prove that the fair
runs of an implementation map to valid runs of a specification. The first set
of proof reductions we present refresh similar attempts in past
work~\cite{RaySumners,cav1} in transferring these proof requirements on
infinite runs to properties about single steps of two systems {\tt impl} and
{\tt spec}. The difference between these past efforts and the work presented is
that we directly specify properties related to guaranteeing progress for each
task in the system and we leverage the definition of the blocking relation. In
addition, while the proof reduction to single step presented in this section
could be used as is, the design of the reduction is influenced by the needs of
subsequent proof reductions over tasks presented in
Section~\ref{sec:trans}. The book ``general-theory.lisp'' in the supporting
materials covers the work in this section.

The goal is to show that if one were to prove certain properties about steps of
an implementation system {\tt impl} and a specification system {\tt spec}, then
one could infer a {\em fair stuttering refinement} -- every fair run of {\tt
  impl} maps to a valid run of {\tt spec}. We wish to prove this for any
specification and implementation system, so specifically, for any {\tt impl}
and {\tt spec} and any fair run {\tt impl-run} of the implementation, if we
have proven the required properties then we can map {\tt impl-run} to a valid
run {\tt spec-run} of {\tt spec}. An overview of the structure of the book
``general-theory.lisp'' is provided in Figure~\ref{fig:general} and attempts to
codify this goal. The definitions of the {\tt impl} and {\tt spec} systems and
the fair run {\tt impl-run} of {\tt impl} are constrained within an encapsulate
to only have the properties: {\tt (def-inf-run impl)}, {\tt (def-fair-pick impl
  id-p)}, {\tt (def-system-props impl id-p)}, {\tt (def-valid-system impl
  id-p)}, and {\tt (def-match-systems impl spec id-p)}. From this fair run {\tt
  impl-run} and the properties proven on {\tt spec} and {\tt impl}, we can
build a valid run {\tt spec-run}. While it is not possible to make this a
closed-form statement of correctness in ACL2, we believe the structure of the
book is sufficient to establish the claim.

The function {\tt (spec-run i)} in Figure~\ref{fig:general} defines the {\tt
  spec} state at each time to simply be {\tt (impl-map (impl-run i))} and the
function {\tt (spec-pick i)} is simply {\tt (impl-pick i)} except that we
introduce finite stutter (i.e. return {\tt nil}) if the mapped state doesn't
change. It is customary to define some notion of observation or labeling of
states that must be preserved to ensure correlation of behavior between {\tt
  spec} and {\tt impl} -- we assume human review has ensured that the mapping
from {\tt impl} states to {\tt spec} states preserves any observations relevant
to the specification. In this regard, it is relevant that the mapped run on the
{\tt spec} is relatively simple in definition as it avoids errors or oversights
in specification due to an obfuscation of how the implementation and
specification are correlated.

\begin{figure}
\scriptsize
\begin{verbatim}
(encapsulate
 (....) ;; constrained functions defining impl and spec.
 ....   ;; local def.s and prop.s to show constraints.
             
;; ASSUMPTIONS:
   ;; assume relevant properties of given systems impl and spec:
 (def-system-props  impl id-p)
 (def-valid-system  impl id-p)
 (def-match-systems impl spec id-p)
   ;; assume an infinite run of the impl system:
 (def-inf-run       impl)
 (def-fair-pick     impl id-p)
)

.... ;; def.s and theorems to establish results.

;; Define the corresponding (assumed to preserve "observations") spec run:
(defun spec-run (i) (impl-map (impl-run i)))

;; spec-pick will introduce stutter into spec-run when the mapped state doesn't change:
(defun spec-pick (i)
  (and (not (equal (impl-map (impl-run (1- i)))
                   (impl-map (impl-run i))))
       (impl-pick i)))

.... ;; additional def.s and theorems to establish results.

;; CONCLUSIONS:
   ;; and prove that the corresponding spec-run is indeed a valid run of spec:
(def-inf-run   spec)
(def-valid-run spec id-p)
\end{verbatim}
\normalsize
\caption{Structure of the book {\tt ``general-theory.lisp''}}
\label{fig:general}
\end{figure}

\begin{figure}
\footnotesize
\begin{itemize}
\item IMPL system definition:
  \begin{itemize}
  \item {\tt (impl-init x)} -- initial predicate on states {\tt x} for impl system
  \item {\tt (impl-next x y k)} -- state {\tt x} transitions to state {\tt y} on selector {\tt k}
  \item {\tt (impl-blok x k)} -- state {\tt x} blocked for transitions for selector {\tt k}
  \end{itemize}
\item SPEC system definition:
  \begin{itemize}
  \item {\tt (spec-init x)} -- initial predicate on states {\tt x} for spec system
  \item {\tt (spec-next x y k)} -- state {\tt x} transitions to state {\tt y} on selector {\tt k}
  \item {\tt (spec-blok x k)} -- state {\tt x} blocked for transitions for selector {\tt k}
  \end{itemize}
\item Definitions needed for {\tt (def-system-props impl id-p)} macro:
  \begin{itemize}
  \item {\tt (impl-iinv x)} -- inductive invariant for states in impl
  \end{itemize}
\item Definitions needed for {\tt (def-match-systems impl spec id-p)} macro:
  \begin{itemize}
  \item {\tt (impl-map x)} -- maps impl states to corresponding spec states
  \item {\tt (impl-rank k x)} -- ordinal decreases until spec matches transition for {\tt k}
  \end{itemize}
\item Definitions needed for {\tt (def-valid-system impl id-p)} macro:
  \begin{itemize}
  \item {\tt (impl-noblk k x)} -- is task id {\tt k} invariantly unblocked in state {\tt x}
  \item {\tt (impl-nstrv k x)} -- ordinal decreases until {\tt k} is in a {\tt noblk} state
  \item {\tt (impl-starver k x)} -- potential starver of {\tt k} in {\tt x} which is not blocked
  \end{itemize}
\end{itemize}
\normalsize
\caption{Function Definitions for Single-Step System-Level Properties}
\label{fig:singdefs}
\end{figure}

The properties we need to prove for {\tt impl} and {\tt spec} are defined by
the macros {\tt def-system-props}, {\tt def-valid-system}, and {\tt
  def-match-systems}. Along with the functions defining the impl and spec
systems, additional definitions are required for each of these macros. We will
shortly go into greater detail on the properties we will assume as constraints
for these functions, but first, we refer to the listing provided in
Figure~\ref{fig:singdefs}.

The macro {\tt (def-system-props impl id-p)} expands into simple theorems
ensuring {\tt (not (id-p nil))}, ensuring {\tt (impl-next x x k)}
is not valid, and ensuring the state predicate {\tt (impl-iinv x)} is an
inductive invariant for {\tt impl} -- namely that {\tt (impl-iinv x)} holds in
the initial state and persists across {\tt (impl-next x y k)} transitions.

The {\tt (def-match-systems impl spec id-p)} macro requires defining {\tt (impl-map x)}, 
a mapping from impl states to spec states and a ranking
function {\tt (impl-rank k x)} which returns an ordinal for each task id {\tt
  k}. The main properties generated by {\tt def-match-systems} are the
following:

\footnotesize
\begin{verbatim}
(defthm map-matches-next
  (implies (and (impl-iinv x) (id-p k) (!= (impl-map x) (impl-map y))
                (impl-next x y k)
                (not (impl-blok x k)))
           (and (spec-next (impl-map x) (impl-map y) k)
                (not (spec-blok (impl-map x) k)))))
\end{verbatim}

\begin{Verbatim}[samepage=true]
(defthm map-finite-stutter
  (implies (and (impl-iinv x) (id-p k) (= (impl-map x) (impl-map y))
                (impl-next x y k))
           (o<  (impl-rank k y) (impl-rank k x))))
\end{Verbatim}

\begin{verbatim}
(defthm map-rank-stable
  (implies (and (impl-iinv x) (id-p k) (id-p l) (!= k l)
                (impl-next x y l))
           (o<= (impl-rank k y) (impl-rank k x))))
\end{verbatim}
\normalsize

The theorem {\tt map-matches-next} ensures that on any step {\tt (impl-next x y
  k)} for task {\tt k} which is not blocked in {\tt x} and where the mapped
specification state changes (i.e. {\tt (!= (impl-map x) (impl-map y))}) then
the spec must be able to match the transition and the spec state cannot be
blocked in the spec for task {\tt k}. The theorem {\tt map-finite-stutter}
ensures that when the mapped implementation state does not change on an update
for task {\tt k} in impl, then the ordinal returned by {\tt impl-rank} must
strictly decrease and the theorem {\tt map-rank-stable} ensures that this
ordinal does not increase when task {\tt k} is not selected. The clear intent
of these properties is to ensure that as long as a task {\tt k} is not
indefinitely blocked when it is selected for update in impl, then eventually a
matching spec transition must be generated. The question is then naturally how
to ensure that a task is not indefinitely blocked. This concept of being
indefinitely blocked is commonly called ``starvation'' in the literature and
the {\tt def-valid-system} macro will generate properties intended to ensure
that no task is starved.

The {\tt (def-valid-system impl id-p)} macro requires the definition of a
predicate {\tt (impl-noblk k x)} which is true when the task {\tt k} can no
longer be blocked in state {\tt x} and a function {\tt (impl-nstrv k x)} which
nominally returns an ordinal that decreases until {\tt (impl-noblk k x)} is
true. Once a task {\tt k} reaches an {\tt impl-noblk} state, it can no longer
be blocked until it transitions and thus the fair selection of {\tt k} will
ensure a transition of {\tt k} occurs. Unfortunately, a task's progress to an
{\tt impl-noblk} state may be dependent on any number of other tasks or
components in the impl state. At this general level of system definition, we
only have system states {\tt x} and task ids {\tt k}, so we imagine that for
any {\tt k} and {\tt x}, we could define a set of task ids called the {\em
  starve-set} which need to make progress before {\tt k} can reach a {\tt
  noblk} state. Updates to ids which are not in this starve-set should simply
have no effect on this progress and so we will assume that {\tt (impl-nstrv k
  x)} will strictly decrease on transitions for ids in the starve-set and
remain unchanged otherwise. Unfortunately, it might be possible that all of the
tasks in the {\em starve-set} are blocked and so we need the additional
definition of an {\tt (impl-starver k x)} which returns an id in this {\em
  starve-set} which is currently not blocked in state {\tt x}. Additionally, we
need to ensure that when an element outside of the starve-set is chosen, that
the {\tt (impl-starver k x)} remains unchanged. The encoding of these
properties as ACL2 theorems are generated from the {\tt def-valid-system} macro
and are listed here:

\scriptsize
\begin{verbatim}
(defthm noblk-blk-thm
  (implies (and (iinv x) (id-p k)
                (noblk k x))
           (not (blok x k))))

(defthm noblk-inv-thm
  (implies (and (iinv x) (id-p k) (id-p l) (!= k l)
                (next x y l)
                (noblk k x))
           (noblk k y)))

(defthm starver-thm
  (implies (and (iinv x) (id-p k)
                (not (noblk k x)))
           (not (blok x (starver k x)))))

(defthm nstrv-decreases
  (implies (and (iinv x) (id-p k) (!= k (starver k x))
                (next x y (starver k x))
                (not (noblk k x)))
           (o< (nstrv k y) (nstrv k x))))

(defthm nstrv-holds
  (implies (and (iinv x) (id-p k) (id-p l) (!= k l)
                (next x y l)
                (not (noblk k x)))
           (o<= (nstrv k y) (nstrv k x))))

(defthm starver-persists
  (implies (and (iinv x) (id-p k) (id-p l) (!= k l) (!= l (starver k x))
                (next x y l)
                (not (noblk k x))
                (= (nstrv k y) (nstrv k x)))
           (= (starver k y) (starver k x))))
\end{verbatim}
\normalsize

And with these properties assumed as constraints, we return to the goal of
proving that the infinite run defined by {\tt (spec-run i)} and {\tt (spec-pick
  i)} from Figure~\ref{fig:general} is indeed a valid run of {\tt spec}. In
order to do that we need to define a function {\tt spec-prog} which satisfies
the requirements set out in Figure~\ref{fig:prog}. First, it is useful to
define an {\tt (impl-prog k i)} and show that the {\tt impl-run} is indeed a
valid run.

The definition of {\tt (impl-prog k i)} is in Figure~\ref{fig:measure} and
essentially looks forward into {\tt impl-run} until we reach an {\tt i}
where {\tt k} is picked and the state changes. The key point is obviously the
question of what is the measure for demonstrating that this function terminates
and this follows from our earlier discussion about the {\tt (impl-noblk k x)},
{\tt (impl-nstrv k x)}, and {\tt (impl-starver k x)} functions. If we have {\tt
  (impl-noblk k ..)} at the current state, then the task with id {\tt k} cannot
be blocked and we can simply countdown the {\tt (impl-fair k i)} measure until
task {\tt k} is selected -- the state will change at that time since {\tt k}
will still be unblocked and {\tt impl-next} must change the state. If {\tt
  (impl-noblk k ..)} does not currently hold then we know there is a task id
{\tt (impl-starver k ..)} which cannot be blocked in the current state and
either {\tt (impl-nstrv k ..)} strictly decreases or {\tt (impl-starver k ..)}
will not change. Thus, at each step, either the {\tt impl-nstrv} measure
strictly decreases or the fair measure for {\tt impl-starver} counts down and
will eventually expire and {\tt impl-nstrv} will strictly decrease.

This {\tt (impl-prog k i)} thus ensures that task {\tt k} is picked and changes
state in {\tt (impl-run i)} but we now must guarantee that the mapped state
changes in {\tt spec}. In the case that the mapped state doesn't change, we
know that the {\tt (impl-rank k ..)} must decrease and that the impl-rank
remains unchanged when other ids are selected. This is the basis for the
definition {\tt (spec-prog k i)} in Figure~\ref{fig:measure}.

\begin{figure}
\scriptsize
\begin{verbatim}
(defun ord-nat-pair (o n)
  ;; simple function which returns lex. product of an o-p o and natp n:
  (make-ord (if (atom o) (1+ o) o) 1 n))

;; First prove that the implementation run is a valid run...
(defun impl-prog (k i)
  (declare (xargs :measure
                  (if (impl-noblk k (impl-run i))
                      (impl-fair k i)
                    (ord-nat-pair (impl-nstrv k (impl-run i))
                                  (impl-fair (impl-starver k (impl-run i)) i)))))
  (cond
   ((or (not (and (natp i) (id-p k)))              ;; ill-formed inputs.. or
        (and (= (impl-pick (1+ i)) k)              ;;    impl-pick matches k
             (!= (impl-run (1+ i)) (impl-run i)))) ;;  ..and k makes progress
    0)
   (t (1+ (impl-prog k (1+ i))))))

;; ...And use that to show that the mapped spec run is also valid
(defun spec-prog (k i)
  (declare (xargs :measure
                  (ord-nat-pair (impl-rank k (impl-run i))
                                (impl-prog k i))))
  (cond
   ((or (not (and (natp i) (id-p k)))              ;; ill-formed inputs.. or
        (and (= (spec-pick (1+ i)) k)              ;;    spec-pick matches k
             (!= (spec-run (1+ i)) (spec-run i)))) ;;  ..and k makes progress
    0)
   (t (1+ (spec-prog k (1+ i))))))
\end{verbatim}
\normalsize
\caption{Defined Measure Functions on Infinite Runs}
\label{fig:measure}
\end{figure}

\section{Proof Reduction to a Small Bounded Number of Tasks} \label{sec:trans}

In the previous section, we presented a proof reduction of the requirements for
fair stuttering refinement from reasoning about infinite runs of systems to
reasoning about single steps of systems. We did not make any assumption about
the state structure of the systems other than that updates occurred
asynchronously at some prescribed task level. In this section, we will assume a
structure on the states of a system and show how to reduce the requisite
properties from across the large state structure to the properties on
components of the state. Throughout this section and the next, we will use the
set {\tt (s k v r)} and get {\tt (g k r)} operations from the records
book~\cite{kaufsum}. In particular, {\tt (g k r)} takes a record {\tt r} and
returns either the value previously set for key {\tt k} in record {\tt r} or
{\tt nil} as default.

\begin{figure}
\footnotesize
\begin{itemize}
\item TR-IMPL system definition:
  \begin{itemize}
  \item {\tt (tr-impl-t-init a k)} -- initial state predicate for t-state {\tt a} and key {\tt k}
  \item {\tt (tr-impl-t-next a b x)} -- t-state {\tt a} transitions to t-state {\tt b} in state {\tt x}
  \item {\tt (tr-impl-t-blok a b)} -- t-state {\tt a} is blocked from stepping by t-state {\tt b}
  \end{itemize}
\item TR-SPEC system definition:
  \begin{itemize}
  \item {\tt (tr-spec-t-init a k)} -- initial state predicate for t-state {\tt a} and key {\tt k}
  \item {\tt (tr-spec-t-next a b x)} -- t-state {\tt a} transitions to t-state {\tt b} in state {\tt x}
  \item {\tt (tr-spec-t-blok a b)} -- t-state {\tt a} is blocked from stepping by t-state {\tt b}
  \end{itemize}
\item Definitions needed for {\tt (def-tr-system-props tr-impl)} macro:
  \begin{itemize}
  \item {\tt (tr-impl-iinv x)} -- inductive invariant as previously.. no change at task-level
  \end{itemize}
\item Definitions needed for {\tt (def-match-tr-systems tr-impl tr-spec)} macro:
  \begin{itemize}
  \item {\tt (tr-impl-t-map a)} -- maps tr-impl t-states to corresponding tr-spec t-states
  \item {\tt (tr-impl-t-rank a)} -- ordinal decreases until mapped t-state must change
  \end{itemize}
\item Definitions needed for {\tt (def-valid-tr-system tr-impl)} macro:
  \begin{itemize}
  \item {\tt (tr-impl-t-noblk a b)} -- is t-state {\tt a} invariantly not-blocked by t-state {\tt b}
  \item {\tt (tr-impl-t-nstrv a b)} -- positive natural which strictly decreases until {\tt (t-noblk a b)}
  \item {\tt (tr-impl-t-nlock k x)} -- ordinal strictly decreases on from {\tt k} to blocker of {\tt k} in {\tt x}
  \end{itemize}
\end{itemize}
\normalsize
\caption{Function Definitions for Single-Step Task-Level Properties}
\label{fig:taskdefs}
\end{figure}

The book ``trans-theory.lisp'' in the supporting materials for this paper
includes the definitions and proofs relating to this section. The structure of
this book is similar to that shown for ``general-theory.lisp'' in
Figure~\ref{fig:general} in that there is an encapsulation which entails the
system definitions and properties we want to assume and then outside of the
encapsulation, we prove the derived results. For the previous section, in
``general-theory.lisp'', we proved the property in Figure~\ref{fig:files} (in 
an abuse of notation pretending ACL2 were higher-order for a moment),
For this section, our goal is to define systems at a task level and derive the
system-level results. In the same higher-level-abuse format as above, we have
the property from ``trans-theory.lisp'' also in Figure~\ref{fig:files}.

\begin{figure}
\scriptsize
\begin{verbatim}
"general-theory.lisp":
  (implies (and (def-system-props  impl id-p)
                (def-valid-system  impl id-p)
                (def-match-systems impl spec id-p))
           (implies (and (def-inf-run   impl)
                         (def-fair-pick impl id-p))
                    (and (def-inf-run   spec)
                         (def-valid-run spec id-p))))
                         
"trans-theory.lisp":
  (implies (and (def-tr-system-props  tr-impl)
                (def-valid-tr-system  tr-impl)
                (def-match-tr-systems tr-impl tr-spec))
           (and (def-system-props  tr-impl key-p)
                (def-valid-system  tr-impl key-p)
                (def-match-systems tr-impl tr-spec key-p)))
\end{verbatim}
\normalsize
\caption{High-Level properties in for theory files definitions}
\label{fig:files}
\end{figure}

We take the state of the system to be a record associating keys to task
states.. what we call {\em t-states}. The task id selected on input is now
simply one of these keys and the update of the state will only update the
corresponding entry of the record. We presume and constrain a fixed finite set
of keys -- {\tt (keys)} -- of arbitrary size and composition and membership in
this set will define the {\tt id-p} test for task id selection. The state of
the system is then a record mapping members of this finite set {\tt (keys)} to
t-states and the system will be defined on the task level. We define task-based
systems by assuming the pertinent definitions on task states in the system and
derive the system-level definitions across the state. We name these systems
derived from the task-level definitions as {\tt tr-impl} and {\tt tr-spec}. In
Figure~\ref{fig:singdefs} from the previous section, we listed the function
definitions required for the single-step system-level properties -- we do the
same for the single-step task-level properties in Figure~\ref{fig:taskdefs}.

Many of the system-level derived functions follow simply from the
task-level. The system-level {\tt (tr-impl-init x)} predicate checks that {\tt
  (tr-impl-t-init (g k x) k)} holds for all keys {\tt k}. The system-level {\tt
  (tr-impl-next x y k)} only updates {\tt (g k x)} as {\tt (tr-impl-t-next (g k
  x) (g k y) x)} and leaves all other keys untouched in {\tt x}. The
system-level block function {\tt (tr-impl-blok x k)} checks if there is any key
{\tt l} such that {\tt (tr-impl-t-blok (g k x) (g l x))}. The system-level
mapping function simply goes through all keys and calls {\tt tr-impl-t-map} for
the corresponding t-state and the system level rank just calls {\tt
  (tr-impl-t-rank (g k x))} directly. The inductive invariant does not change;
there is just one inductive invariant defined on the entire record defining the
system state. Additionally, the system-level proofs for {\tt (def-system-props
  tr-impl key-p)} and {\tt (def-match-systems tr-impl tr-spec key-p)} are
straightforward and follow from these system-level definitions and properties
of task-level definitions.

The functions and properties for proving progress and valid {\tt impl} runs are
more involved. For the sake of brevity and readability, we will drop the {\tt
  tr-impl-} prefix from the system-level and task-level defintions for the
remainder of this section. In addition to ensuring that {\tt t-nlock} returns
an ordinal and {\tt t-nstrv} returns a positive natural
number\begin{footnote}{In the supporting materials for this paper, {\tt
    t-nstrv} is generalized to be a list of natural numbers which is then
  summed and combined into a list of lists of naturals, but for the sake of
  clarity and brevity in this paper, we keep a simpler definition for {\tt
    t-nstrv}. We could not use a generic ACL2 ordinal for {\tt t-nstrv} since
  we needed to form lexicographic products of sums of these ordinals and that
  is not possible for arbitrary ordinals in ACL2.}\end{footnote}, the macro
{\tt (def-valid-tr-system tr-impl)} introduces the following properties:

\scriptsize
\begin{verbatim}
(defthm t-noblk-blk-thm
  (implies (and (iinv x) (key-p k) (key-p l)
                (t-noblk (g k x) (g l x)))
           (not (t-blok (g k x) (g l x)))))

(defthm t-noblk-inv-thm
  (implies (and (iinv x) (key-p k) (key-p l)
                (t-noblk (g k x) (g l x))
                (t-next (g l x) c x))
           (t-noblk (g k x) c)))

(defthm t-nlock-decreases
  (implies (and (iinv x) (key-p k) (key-p l)
                (t-blok (g k x) (g l x)))
           (o< (t-nlock l x)
               (t-nlock k x))))

(defthm t-nstrv-decreases
  (implies (and (iinv x) (key-p k) (key-p l)
                (not (t-noblk (g k x) (g l x)))
                (not (t-noblk (g k x) c))
                (t-next (g l x) c x))
           (< (t-nstrv (g k x) c)
              (t-nstrv (g k x) (g l x)))))
\end{verbatim}
\normalsize

The system-level {\tt (noblk k x)} definition simply checks that {\tt (t-noblk
  (g k x) (g l x))} holds for every key {\tt l} and as such, the task-level
{\tt t-noblk-blk-thm} and {\tt t-noblk-inv-thm} are task-level projections of
their system-level counterparts and the system-level properties follow fairly
easily. The more interesting case comes up in defining the system-level {\tt
  (nstrv k x)} and {\tt (starver k x)}.  For the task-level, the property {\tt
  t-nlock-decreases} ensures that we don't have any ``deadlocks'' or simply
that for any set of keys, there is always some key in that set which is not
blocked in {\tt x} by some other key in that set. The combination of {\tt
  t-nstrv-decreases} and the properties of {\tt t-noblk} ensure that no task
can be starved by another task.

The intuition behind defining the system-level {\tt (nstrv k x)} begins by
recognizing that if {\tt (not (noblk k x))} then there is some set of keys {\tt
  l} such that {\tt (not (t-noblk (g k x) (g l x)))}. We will call this set of
keys the {\em may-block set}. But since {\tt t-noblk} persists once we reach
it, then we could sum up the {\tt (t-nstrv (g k x) (g l x))} for this may-block
set and the resulting ordinal would decrease until we reached a state where
{\tt k} was {\tt t-noblk} for all {\tt l} and thus {\tt noblk}. Assume for the
moment that {\tt k} were not blocked (i.e. we could set {\tt (starver k x)} to
be {\tt k}), then consider an update for some key {\tt l}. If that key were in
the may-block set of {\tt k} then the ordinal would decrease. If {\tt l} is not
in the may-block set of {\tt k} then {\tt (t-noblk (g k x) (g l x))} and the
transition of {\tt l} cannot change the blocked status of {\tt k} and it cannot
change the may-block set for {\tt k} and so progress is made. Unfortunately
there is no guarantee that {\tt k} is not blocked and thus we cannot pick a
suitable {\tt starver} which ensures progress when selected.

But from the property {\tt t-nlock-decreases}, starting with {\tt k} in {\tt
  x}, we can find a key which is not blocked by checking if the key is blocked
and recurring on the first blocking key we find if we are blocked. This is the
definition of the function {\tt (starver k x)} and is included here:

\small
\begin{verbatim}
(defun starver (k x)
  (declare (xargs :measure (t-nlock (g k x))))
  (if (and (iinv x) (key-p k) (blok x k))
      (starver (pikblk k x) x) 
    k))
\end{verbatim}
\normalsize

The function {\tt (pikblk k x)} returns the first key we find such that {\tt
  (t-blok (g k x) (g (pikblk k x)))}. So, from {\tt k}, we can find a key which
is unblocked, but the question is then how to build a measure from the
starve-set including {\tt k} and {\tt (starver k x)}. The answer is to build a
natural list where each element is the sum of {\tt t-nstrv} for the may-block
set (as we described before) in each step along the path from {\tt k} to {\tt
  (starver k x)} and define our ordinal as the lexicographic product of the
naturals in this list. The first observation is that at the end of this list we
will have the summation of {\tt t-nstrv}s for the may-block set of {\tt
  (starver k x)} and since {\tt (starver k x)} is not blocked, it will make
progress as we discussed before. The other key observation is that at each
step, the {\tt (pikblk k x)} key will be in the may-block set of {\tt k} and
thus even though a transition of {\tt (pikblk k x)} may modify its may-block
set and potentially increase the measure from that point, the measure for the
may-block set of {\tt k} will decrease and the ordinal over all will
decrease. This list of naturals is defined by the function {\tt (nstrvs* k x)}
as follows where the function {\tt (scar s)} and {\tt (scdr s)} return the
first element and remainder of a set respectively and {\tt (card s)} returns
the cardinality of the set.

\footnotesize
\begin{Verbatim}[samepage=true]
(defun sum-nsts* (k x s)
  (declare (xargs :measure (card s)))
  (if (null s) 1
    (+ (if (t-noblk (g k x) (g (scar s) x)) 0
         (t-nstrv (g k x) (g (scar s) x)))
       (sum-nsts* k x (scdr s)))))

(defun sum-nsts (k x) (sum-nsts* k x (keys)))

(defun nstrvs* (k x)
  (declare (xargs :measure (t-nlock (g k x))))
  (if (and (iinv x) (key-p k) (blok x k))
      (cons (sum-nsts k x) (nstrvs* (pikblk k x) x))
    (list (sum-nsts k x))))

(defun nats->o (n l)
  (cond ((zp n) 0)
        ((atom l) (make-ord n 1            (nats->o (1- n) ())))
        (t        (make-ord n (1+ (car l)) (nats->o (1- n) (cdr l))))))

(defun tr-impl-nstrv (k x)
  (nats->o (card (keys)) (nstrvs* k x)))
\end{Verbatim}
\normalsize

As we mentioned, the function {\tt nstrvs*} returns a natural list and we build
a suitable ordinal from this list using the function {\tt nats-o}. But because
the length of the path to {\tt (starver k x)} from {\tt k} could change and
thus the length of the nstrvs* list could change, we need to make the defined
ordinal ``first-aligned'' -- where the first element in the list is mapped to a
coefficient of the same exponent no matter the length of the rest of the
list. We use {\tt (card keys)} as the starting exponent and prove separately
that the length of the list returned by {\tt nstrvs*} can never exceed {\tt
  (card keys)}.

This construction also shows one of the reasons we assume an arbitrary fixed
finite set of {\tt (keys)} (in order to put a bound on {\tt (len (nstrv* k
  x))}), but this restriction makes sense for other reasons as well. If the set
of keys were not finite, then we would need some additional requirement to
ensure that a task were not persistently blocked by an infinite sequence of
newly instantiated tasks. Other options exist to avoid this (such as requiring
that all new tasks cannot block existing tasks) but these alternatives end up
imposing constraints we believe are too restrictive.

\section{Example -- A Bakery Algorithm} \label{sec:bakery}

We use the Bakery algorithm as an example application of the proof reductions
we present in this paper. The Bakery algorithm was developed by
Lamport~\cite{Lamport} as a solution to mutual exclusion with the additional
assurance that every task would eventually gain access to its exclusive
section. The Bakery algorithm has also been a focus of previous ACL2 proof
efforts~\cite{Ray}.

The essential idea of the algorithm is that each task first goes through a
phase where it chooses a number (much like choosing a number in a bakery) and
then later compares the number against the numbers chosen by the other tasks to
determine who should have access to the exclusive section. The version of the
Bakery algorithm we will use is defined in Figure~\ref{fig:bakeimpl} (the {\tt
  (upd r .. updates ..)} simply expands into a nest of record sets).

Each task will start in program location 0 and start its {\tt :choosing}
phase. During the {\tt :choosing} phase, the task will grab the current shared
max (via the function {\tt (curr-sh-max x)}) and then set its own position {\tt
  :pos} to be 1 more than the shared max. In program {\tt :loc} 3, a
compare-and-swap is implemented and the shared-max is potentially updated. The
task then ends its {\tt :choosing} phase.

After the {\tt :choosing} phase, the task will enter program locations 5 and
6. In these locations, the {\tt t-blok} predicate ensures that the task wait
until other tasks are not {\tt :choosing} and then wait until it has the least
position (where potential ties are broken by comparing the {\tt ndx} of the
{\tt :key} in the set {\tt (keys)}.

\begin{figure}
\scriptsize
\begin{verbatim}
(defun bake-impl-t-init (a k)
  (= a (upd nil :loc 0 :key k :pos 1 :old-pos 0 :temp 0 :sh-max 1)))

(defun bake-impl-t-next (a b x)
  (case (g :loc a)
        (0 (= b (upd a :loc 1 :choosing t)))
        (1 (= b (upd a :loc 2 :temp (curr-sh-max x))))
        (2 (= b (upd a :loc 3 :pos (1+ (g :temp a))
                               :old-pos (g :pos a)
                               :pos-valid t)))
        (3 (= b (upd a :loc 4 :sh-max (if (> (curr-sh-max x) (g :temp a))
                                           (curr-sh-max x)
                                         (g :pos a)))))
        (4 (= b (upd a :loc 5 :choosing nil)))
        (5 (= b (upd a :loc 6))) ;; we are potentially blocked here
        (6 (= b (upd a :loc 7))) ;; we are potentially blocked here
        (t (= b (upd a :loc 0 :pos-valid nil)))))

(defun bake-impl-t-blok (a b)
  (or (and (= (g :loc a) 5)
           (g :choosing b))
      (and (= (g :loc a) 6)
           (and (g :pos-valid b)
                (lex< (g :pos b) (ndx (g :key b))
                      (g :pos a) (ndx (g :key a)))))))
\end{verbatim}
\normalsize
\caption{Bakery Implementation System Definition}
\label{fig:bakeimpl}
\end{figure}

In order to prove {\tt (def-valid-tr-system bake-impl)}, we need to define the
{\tt t-nlock}, {\tt t-noblk}, and {\tt t-nstrv} functions. The definition of
{\tt (t-nlock x k)} needs to return an ordinal that is strictly decreasing from
the blocked task to the blocking task. From the {\tt bake-impl-t-blok}
relation, we note that {\tt :choosing} states cannot be blocked and that {\tt
  lex<} is already well-founded, so we can devise a suitable {\tt
  bake-impl-t-nlock}:

\footnotesize
\begin{Verbatim}[samepage=true]
(defun bake-impl-t-nlock (k x)
  (let ((a (g k x)))
    (make-ord 2 (if (g :choosing a) 1 2)
    (make-ord 1 (1+ (nfix (g :pos a)))
                (ndx (g :key a))))))
\end{Verbatim}
\normalsize

For the {\tt t-noblk} and {\tt t-nstrv} definitions, we need to analyze where
one task can no longer block another task. The simple answer is that {\tt
  (t-noblk a b)} is reached once task {\tt b} has chosen a {\tt :pos} greater
than the one in {\tt a}, but we also have to make sure that task {\tt b} is not
choosing either. In addition, we note that if {\tt a} cannot currently be
blocked by any task, then we can set {\tt t-noblk} and task {\tt a} cannot be
blocked if it is not in program locations 5 or 6. With that, we define {\tt
  bake-impl-t-noblk}:

\footnotesize
\begin{verbatim}
(defun bake-impl-t-noblk (a b)
  (or (and (!= (g :loc a) 5)
           (!= (g :loc a) 6))
      (and (not (g :choosing b))
           (> (g :pos b) (g :pos a)))))
\end{verbatim}
\normalsize

Finally, we need to define {\tt t-nstrv} which counts down until we reach the
{\tt t-noblk} state. The simple answer would be to count from the exit of {\tt
  :choosing} phase until the next exit from the {\tt :choosing} phase. Thus, we
would return 8 if {\tt (g :loc b)} was 5 and then proceed down to 6 for 7, then
5 for 0 (wrapping back), then down to 1 for 4 (end of next {\tt
  :choosing}). This almost works.. except that it is possible for {\tt b} to be
in {\tt :loc} 2, 3, or 4 with a {\tt :pos} lower than {\tt a} but {\tt a} has
proceeded further. Thus, we need to add a few steps for the case of being in
2,3,4 with a potentially lower {\tt :pos} but when we come back around for the
next {\tt :choosing}, we will reach {\tt noblk}:

\footnotesize
\begin{verbatim}
(defun bake-impl-t-nstrv (a b)
  (pos-fix
   (cond ((or (and (= (g :loc b) 2)
                   (< (g :temp b) (g :pos a)))
              (and (> (g :loc b) 2)
                   (<= (g :pos b) (g :pos a))))
          (+ 8 (- 8 (g :loc b))))
         ((>= (g :loc b) 5) 
          (+ 5 (- 8 (g :loc b))))
         (t
          (+ 0 (- 5 (g :loc b)))))))
\end{verbatim}
\normalsize

With these definitions and a suitable invariant {\tt bake-impl-iinv}, we can
prove the theorems for (def-valid-tr-system bake-impl) -- each of which just
blasts into a big case split which pushes through. For the specification of the
bakery algorithm, we have a simple system {\tt bake-spec} defined in
Figure~\ref{fig:bakespec}. Each task in this system goes through the following
steps: first, load up a new provisional {\tt :pos} in the {\tt :load} variable,
then proceed to set the {\tt :pos} variable and begin to arbitrate in the {\tt
  'interested} state. Tasks are blocked if some other task is in the {\tt 'go}
state or is in the {\tt 'interested} state and has a lower {\tt :pos}. The
definitions and proof of {\tt (def-match-tr-systems bake-impl bake-spec)} are
fairly straightforward and included in Figure~\ref{fig:bakespec}. We note that
it is feasible (although not required) to define the supporting functions and
prove {\tt (def-valid-tr-system bake-spec)} -- this proves that all fair runs
of {\tt bake-spec} are valid while the earlier proofs only ensured that the
runs mapped from {\tt bake-impl} runs were valid.

In previous work~\cite{RaySumners}, a similar proof effort was conducted in
proving a fair stuttering refinement for the definition of the Bakery
Algorithm. In that effort, the proof was complicated by the need to add
additional structures to track fair scheduling and to ensure correlation to a
specification which had additional structures to ensure progress for each
task. These complications were avoided in the proof here and as such, much less
definition and details were required. The reduced proof we present here is
primarily the definition and proof of a sufficient inductive invariant but much
additional definition and proof was required in the earlier
work~\cite{RaySumners}.

\begin{figure}
\footnotesize
\begin{verbatim}
(defun bake-spec-t-init (a k)
  (declare (ignore k))
  (and (= (g :loc a) 'idle) (= (g :pos a) 0) (= (g :load a) 0)))

(defun bake-spec-t-next (a b x)
  (case (g :loc a)
    (idle       (and (= (g :loc b)  'loaded)
                     (= (g :pos b)  (g :pos a))
                     (natp (g :load b))
                     (>  (g :load b) (max-pos x))
                     (>= (g :load b) (max-load x))))
    (loaded     (= b (upd a :loc 'interested
                             :pos (g :load a))))
    (interested (= b (upd a :loc 'go)))
    (go         (= b (upd a :loc 'idle)))))

(defun bake-spec-t-blok (a b)
  (and (= (g :loc a) 'interested)
       (or (= (g :loc b) 'go)
           (and (= (g :loc b) 'interested)
                (< (g :pos b) (g :pos a))))))

(defun bake-impl-t-map (a)
  (upd nil
       :loc  (case (g :loc a)
                   ((0 1) 'idle)
                   ((2 3) 'loaded)
                   ((4 5 6) 'interested)
                   (t 'go))
       :pos  (case (g :loc a)
                   (3 (g :old-pos a))
                   (t (g :pos a)))
       :load (case (g :loc a)
                   (2 (1+ (g :temp a)))
                   (t (g :pos a)))))

(defun bake-impl-t-rank (a)
  (case (g :loc a)
        (0 1)       (1 0)
        (2 1)       (3 0)
        (4 2) (5 1) (6 0)
        (t 0)))
\end{verbatim}
\normalsize
\caption{Bakery Specification System and Definitions for Proving Matching from Impl}
\label{fig:bakespec}
\end{figure}

\section{Further Reductions and Considerations} \label{sec:further}

We conclude this paper with a discussion of further reductions and
considerations for search procedures. We first acknowledge that some of the
task-based definitions may seem overly restrictive. For example, the {\tt (blok
  a b)} relation being defined simply on task states. In essence, this
restricts us from supporting systems where a task may be blocked when only some
combination of tasks exist. It is possible to extend the notion of blocking to
be more general but it comes at the cost of the complexity of other definitions
and checks and we have generally found that by adding auxiliary variables to
the task state, we can fit any appropriate system under these restrictions.

This paper focused on mechanized proof reductions for general system
definitions, but the work also supports improvements in more efficient
automatic verification (in particular when the underlying task state space is
finite). For example, take a somewhat draconian restriction that {\tt (t-next a
  b x)} can be defined as {\tt (t-next a b)} and similarly, the initial state
predicate ignored the input {\tt k} -- a few things develop in this
case. First, we note (somewhat trivially) that for every reachable system state
composed of (say) $n$ task states, that every ``substate'' of $n-1$ task states
can also be reached. Additionally, if the task state space were finite, then we
could compute all of the potential cycles in the blocking relation and for each
cycle of size $n$, we could determine if it was reachable by searching through
the system states with only $n$ keys. A similar check could be implemented for
the other properties with no more than 2 keys needed.

Of additional interest in this case, is that reachable states of these systems
have a particular characterization. Consider any run of a system.. any steps in
the run can be permuted as long as the permutation does not change the blocking
relationship between the tasks involved. This means that for every reachable
state, one can define a set of canonical runs which involves only stepping
tasks until the blocking relationship is changed with respect to another task
and then switching to the blocking task or stepping back and switching to the
blockee task. This property limits the structure of potential invariants and
suggests procedures for proving invariants over pairs of states. The inductive
invariant {\tt iinv} over the system state can be defined by invariant
definitions on single task states, pairs of states, triples, etc. and in most
cases (potentially with additional auxiliary variables), sufficiently defined
on single t-states and pairs of t-states. In this case, the requisite
properties of the defined {\tt t-nlock}, {\tt t-nstrv}, {\tt t-noblk}, {\tt
  t-map} and {\tt t-rank} definitions could be proven via GL on the specified
finite t-state domain using a SAT solver with a sufficient conditions on the
t-states assumed. An inductive invariant (defined on single t-states and pairs
of t-states) could be defined that proved each of these sufficient condition
assumptions as invariant of the system. A model checker could be used to reduce
the definitional requirements further by checking invariants (not requiring
inductive invariants) and by checking for bad cycles to show that one could
infer the existence of suitable {\tt t-nlock}, {\tt t-nstrv}, and {\tt
  t-rank}. The model checking problems could be limited to a small number of
tasks and possibly only single task stepping depending on the conditions of the
defintion. The work presented in this paper is a step into many potential
future directions.

\nocite{*}
\bibliographystyle{eptcs}
\bibliography{mybib}
\end{document}